\documentclass{kapproc} % Computer Modern font calls
\usepackage{procps} 

\usepackage[dvips]{graphicx}

\upperandlowercase

\newcommand{\be}{\begin{equation}}
\newcommand{\ee}{\end{equation}}
\newcommand{\bea}{\begin{eqnarray}}
\newcommand{\eea}{\end{eqnarray}}

\newcommand{\Rv}{R_{200}}

\newcommand{\BM}{$\beta$-model}
\newcommand{\bm}{\BM\ }

\begin{document}
 
\articletitle[SPH Simulations of Galaxy Clusters]
{Study of galaxy cluster properties from high-resolution SPH  simulations}

\author{G. Yepes$^1$, Y. Ascasibar$^{1,2}$, S. Gottlöber$^3$ and V. Müller$^3$ }
\affil{$^1$  Grupo Astrofísica, Universidad Autónoma de Madrid\\
$^2$ Theoretical Physics, Oxford University \\
$^3$ Astrophysikalisches Institut Potsdam}

%% affil, email, and abstract are optional

%% optional, to supply a shorter version of the title for the running head:
%%\chaptitlerunninghead{}

%\anxx{Beerends\, John G.}

\begin{abstract}

We  present  some of the results  of an ongoing collaboration
to sudy the dynamical properties of galaxy clusters by means of
high resolution adiabatic SPH cosmological simulations. Results from
our  numerical clusters have been  tested against analytical models
often used in X-ray observations:
 $\beta$ model (isothermal and polytropic) and  those based on
 universal dark matter profiles.  We find a universal temperature
 profile, in agreement with  AMR gasdynamical
 simulations of galaxy clusters. Temperature decreases 
by  a factor 2-3  from the center to virial radius.
  Therefore, isothermal models (e.g. $\beta$ model)
 give a very poor fit to simulated data. Moreover, 
  gas entropy profiles   deviate from  a power law near  the
 center, which is also in very good agreement with  
independent  AMR simulations.
  Thus, if future X-ray observations confirm that gas in
 clusters has an extended isothermal core, then non-adiabatic physics 
 would be required in order  to explain it.
\end{abstract}

%\begin{keywords}
% Galaxy Clusters: properties,  Numerical simulations, Cosmology 
%\end{keywords}

\section{Introduction}
Clusters of galaxies are the largest gravitationally bound structures in the
universe. Therefore,  they have often been considered as a canonical
data set for cosmological tests.
During the last two decades, a great effort has been devoted to
investigate the mass distribution in  CDM haloes by means of numerical
N-body simulations. It is now  firmly established that 
dark matter  density profiles can be fitted by an universal 
 two-parameter function, valid from galactic to cluster scales.
For  the gas component, the situation is  less clear.
The ICM  is in the form of a hot 
diffuse X-ray emitting plasma, where the cooling time
(except in the innermost regions) is typically
longer than the age of the universe.
Adiabatic gasdynamical simulations have therefore been used to study
the formation and evolution of galaxy groups and clusters in different 
cosmologies. 
The Santa Barbara Cluster Comparison Project (\cite{SB_sh}, SBCCP)
 showed a clear 
difference between SPH and Eulerian Adaptive Mesh Refinement (AMR) codes.
 While (the only one available at that time)  Bryan and Norman's  AMR
 code   predicted  an isentropic  gas profile  at the center,  
all the SPH codes used  in SBCCP   predicted an
isothermal gas distribution  almost to the  virial radius of the Coma-like
simulated cluster.  As  pointed out by several people
(e.g. \cite{Lewis00_sh}; \cite{deva}), 
 the standard SPH method could suffer from entropy conservation
 problems. This is particularly  accentuated in low mass resolution SPH
 simulations (see \cite{Borgani02_sh}).
   A new implementation of SPH has been recently proposed
(\cite{gadgetEntro02}) in which  entropy conservation is 
 much better fullfilled. 

In order to asses the reliability of our numerical results, we did 
 an extensive convergence study in terms of  resolution (mass and
spatial), as well as numerical  technique. For this last purpose, we 
resimulated  one of our  clusters  with 3 different numerical codes:
Tree-SPH GADGET, both with the 
standard  (\cite{gadget01}) and  the  entropy conserving SPH
implementation (\cite{gadgetEntro02}),  as well as  the  Eulerian AMR
code ART (\cite{ARThydro02}). Radial profiles  of   gas
and dark matter  are compared  in Figure 1. 
The agreement between AMR and the entropy version of GADGET is
remarkable. The  standard SPH GADGET  still shows the same trend 
 reported  in Frenk et al. (1999), although our mass resolution is
 64 times better ($512^3$  effective  particles). 

\section{Numerical experiments}

 We have carried out a series of high-resolution
 gasdynamical simulations of cluster formation in a flat LCDM universe 
($\Omega_{\rm m}=0.3$; $\Omega_\Lambda =0.7$; $h=0.7$; $\sigma_8=0.9$; $\Omega_{\rm b}=0.02\
h^{-2}$). Simulations were  run with  the 
entropy conserving SPH version of the parallel Tree code  GADGET. 
%Same  clusters were also run with N-body code  have been also done  with  ART, but with the N-body
%part only, expect for the cluster that is shown in Figure 1 (left) in
%which the hydro-ART was used.
 We  have  selected  15 clusters
extracted from a low-resolution ($128^3$) volumen of   $ 80 h^{-1}$
Mpc.  Each object has been re-simulated by means of
the multiple mass technique (e.g. \cite{Klypin01}).
We use 3 levels of mass refinement, reaching an effective resolution
of $512^3$ CDM particles ($\sim 3 \times  10^8\ h^{-1}$ M$_\odot$). 
Gas has been added in the highest resolved area only. The gravitational
smoothing was set to  $\epsilon=2-5\ h^{-1}$ kpc,
 depending on number of particles within the virial radius
 (\cite{Power03_sh}). The minimum smoothing length for SPH was fixed to
 the same value as $\epsilon$.   The X-ray temperature of these objects ranges
  from 1 to  3 keV.   
For a 
 more extended   description of the numerical experiments, the reader is
 referred to  Ascasibar (2003).

\begin{figure}[t]
   \includegraphics[height=4.5cm]{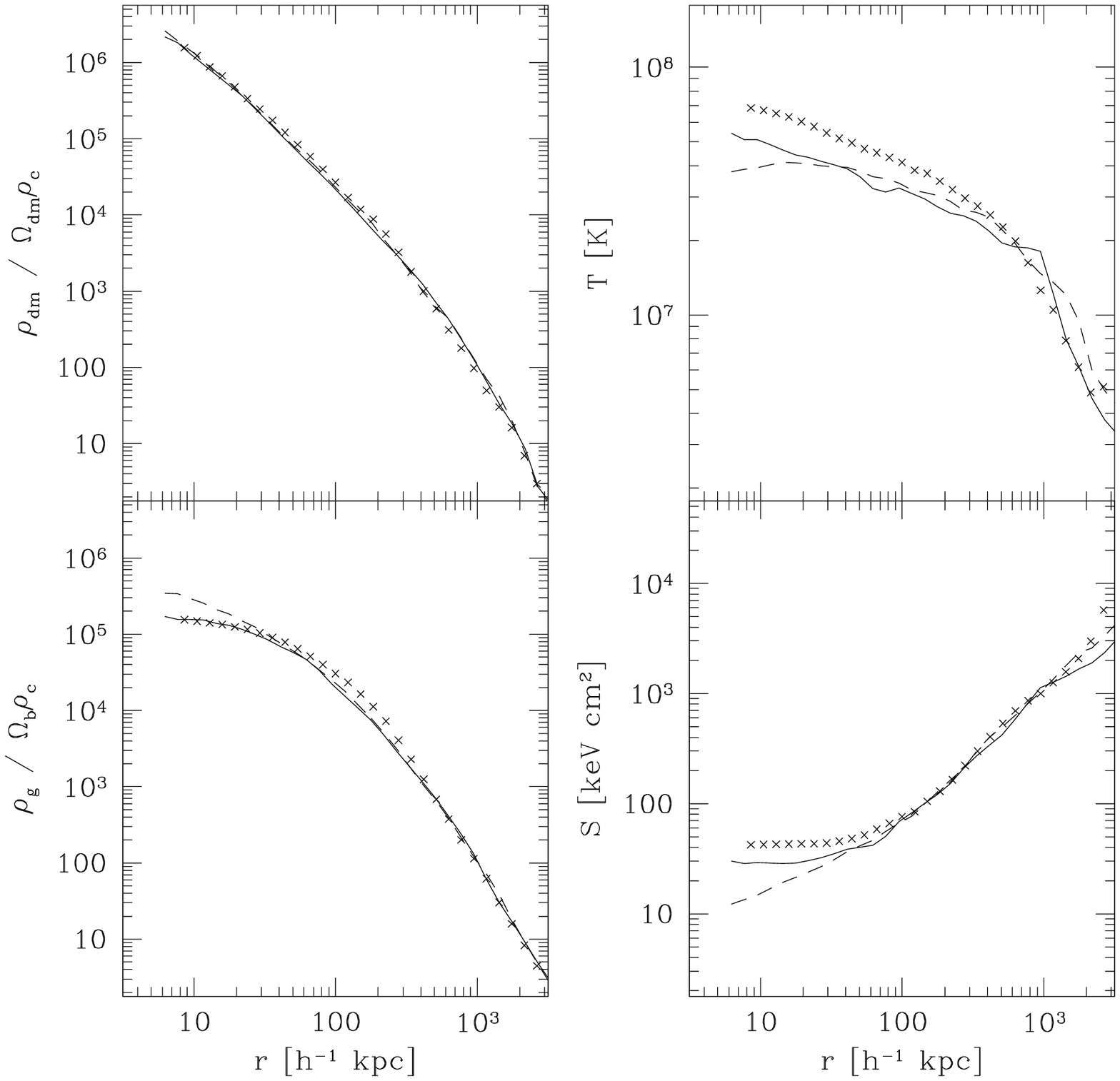} \hspace*{0.6cm} \includegraphics[height=4.5cm]{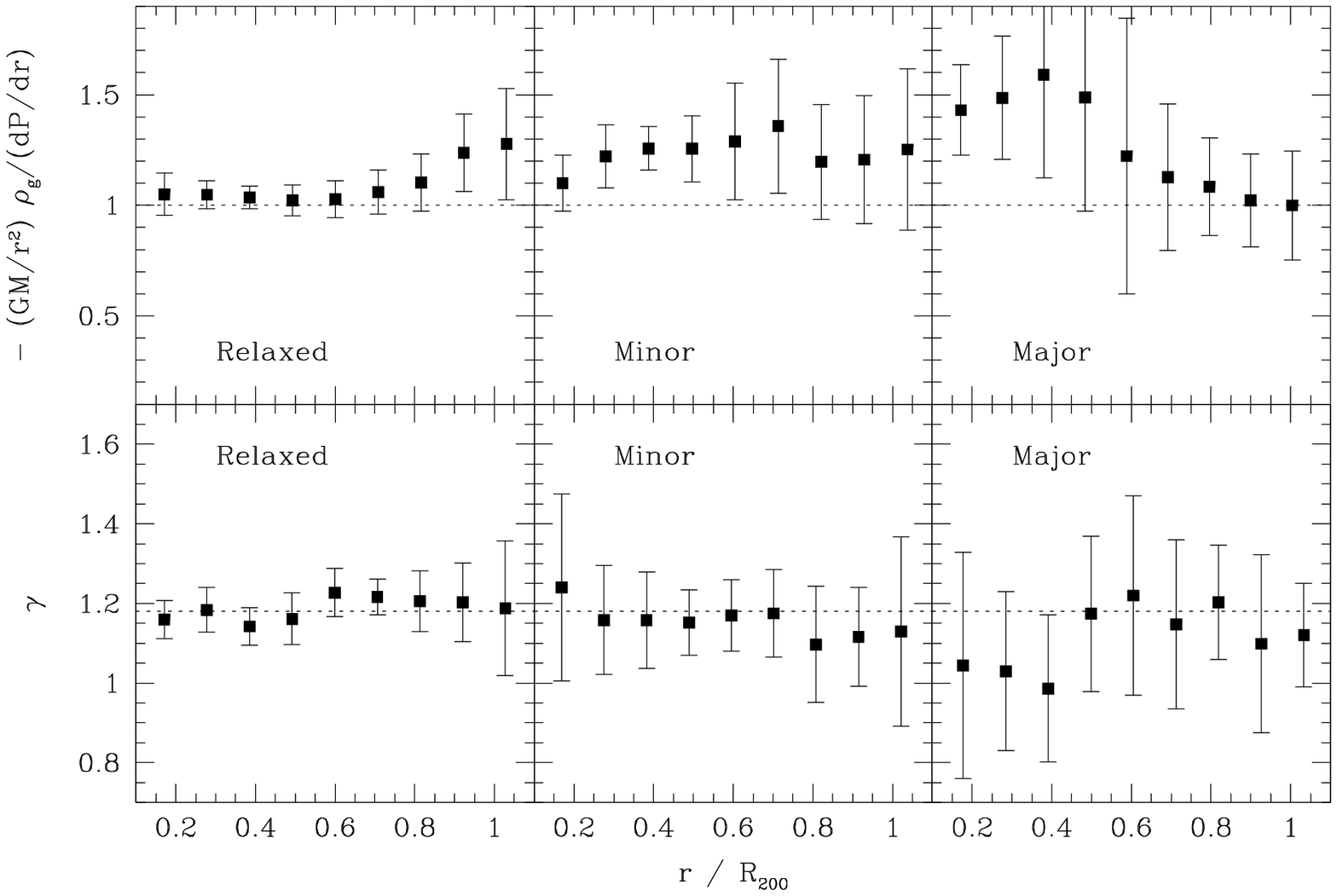}
  \caption{ {\bf Left:} Comparison of density, temperature and entropy profiles for
   a cluster simulated with 3 different numerical hydro codes: 
    Standard SPH GADGET (dashed lines);  Entropy-conserving GADGET 
 (solid lines) and  eulerian AMR ART code 
    (crosses). {\bf Right:} Testing Hydrostatic equilibrium (upper
    panel) and polytropic equation of state  (lower panel) in
    our numerical clusters, classified  according to their dynamical state. }
  \label{figA}
\end{figure}
%__________________________________

\section{Results}

A  detailed   discussion  of the results from our numerical
experiments  can be found elsewhere 
(\cite{tesis_short}; \cite{Ascasibar03}). Here,  we will focus on the
radial structure  of gas and dark matter in  clusters.

We have considered four self-consistent analytical models,
 based on the hypotheses that the hot ICM gas is in 
hydrostatic equilibrium with the dark matter halo and that it follows a
polytropic equation of state.
Two of our models assume NFW (\cite{NFW97})  and MQGSL (\cite{Moore99}) formulae to describe the CDM
density profile, whereas the other two assume a \bm for the gas
distribution. One is an isothermal version with $\beta=2/3$ (BM) and the
other is a polytropic model with $\gamma=1.18$ and $\beta=1$ (PBM).
We have first tested the hypothesis of
hydrostatic equilibrium (HE) and polytropic equation of state (e.o.s) for the gas in
our clusters. In Figure 1 (right) we show the results of this test.  H.E.
is nicely fulfilled by those clusters that are in a  relaxed or  minor
merger state. The  e.o.s for the gas in  these clusters can be reasonably
approximated by a  constant polytropic index of $\gamma \sim 1.2$. 
Then, we compared the simulated radial distributions of gas and dark
matter  with each analytical model. 
By fitting the numerical  X-ray surface brightness,   the 
models based on  universal CDM profiles are able to estimate
    the ICM properties within $30-40\%$  errors.     $\beta$-models
  yield similar estimates  for $r\geq0.1\Rv$, but the shape of the
  inferred profiles at    smaller radii are  severely misleading.

In Figure 2  (left) we plot the spherically  averaged  
 temperature  profile of 
our halos, together with the best fit  for  each 
analytical model. As can be seen, the  $\beta$-model gives
the poorest fit because  gas in simulations  is far from being isothermal. 

The projected emission-weighted temperature profile 
is also shown in Figure 2 (right), 
compared with recent  AMR cluster simulations (\cite{Loken02_sh}) and
with  X-ray observations.   Both sets of simulations
 predict the same   universal  gas temperature  profile for clusters, 
 with no indication of an isothermal core. This is one of the few
   cases in which  SPH and AMR simulations agree so well on one issue.
If the existence of a large isothermal core  is indeed confirmed by 
upcoming   X-ray observations, it would be an indication that 
 non-adiabatic processes must be considered in numerical simulations of
 galaxy cluster formation.

%__________________________________
\begin{figure}
  \centering \includegraphics[height=5cm]{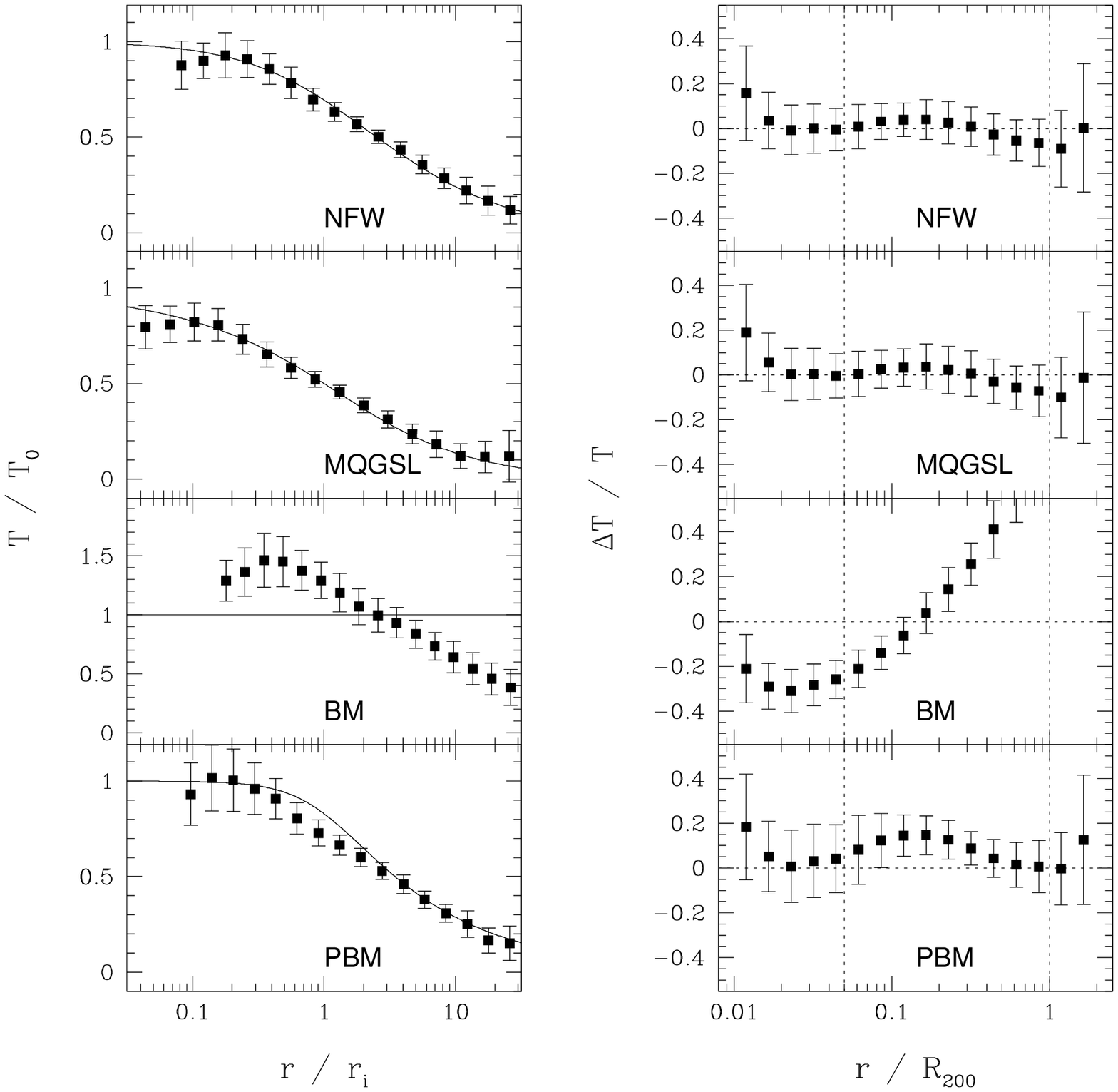}
  \hspace*{0.5cm} \includegraphics[width=5cm]{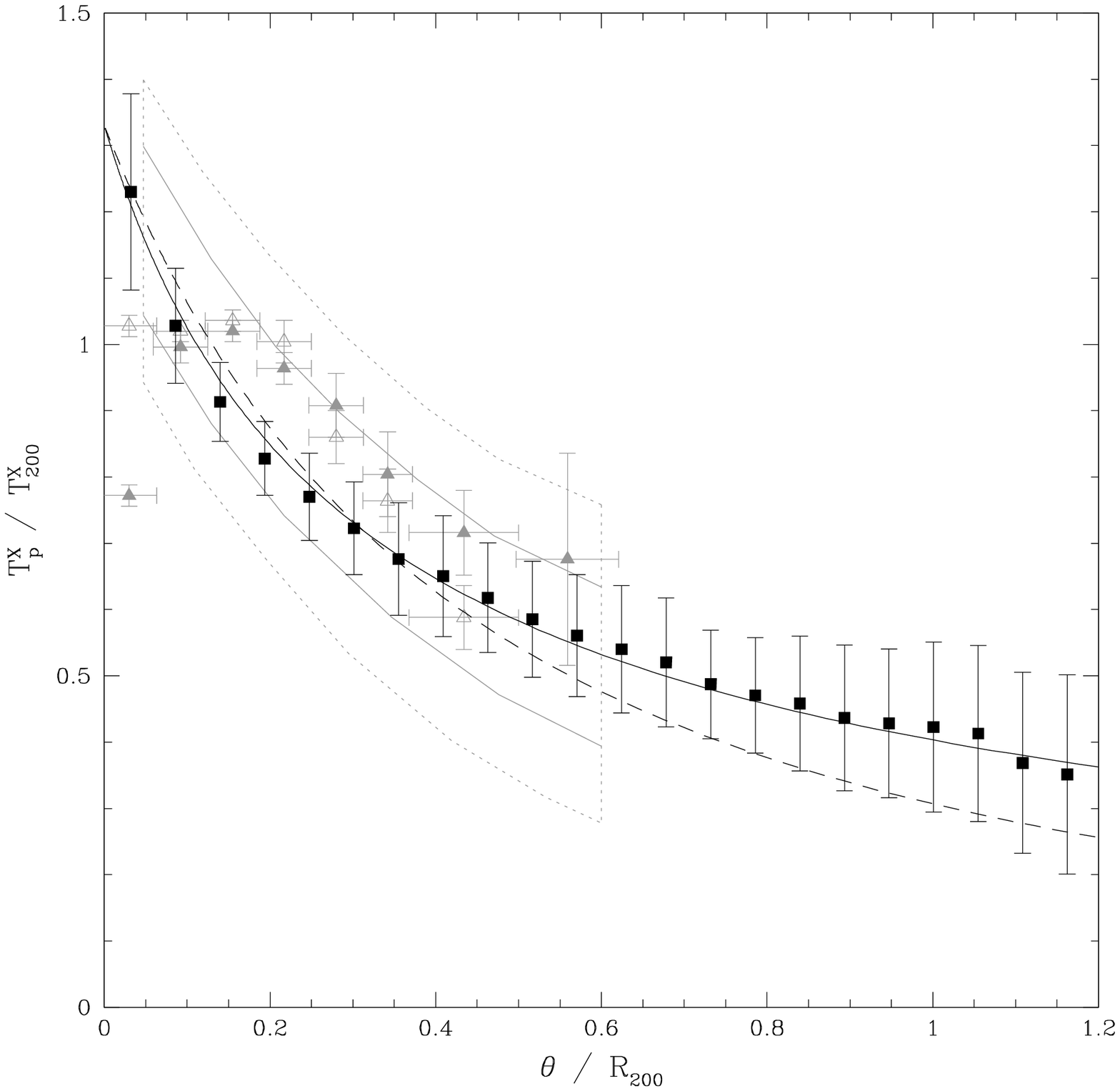}
  \caption{
    {\bf Left:} 3d Temperature profiles  from clusters (points) and the
    corresponding fit from 4 analytical models (see \cite{Ascasibar03}
    for more information).
{\bf Right:} Projected emission-weighted temperature profile
    (black squares with  error bars). Dashed  line:  results from   AMR simulations 
    (\cite{Loken02_sh}), solid  line fit to our data
    \cite{Loken02_sh}. Observational data from \cite{GrandiMolendi02}
    (triangles) and \cite{Markevitch98}  (enclosed boxes).}

  \label{figTx}
\end{figure}
%__________________________________

%\section{Conclusions}

\bibliographystyle{kapalike}
\chapbblname{yepes_mikonos}
\chapbibliography{../BIBTEX/DATABASE,../BIBTEX/PREPRINTS,../BIBTEX/SHORT}

%\end{chapthebibliography}

\end{document}